\documentclass[12pt]{article}

\usepackage{graphicx}

\def\ii{\'{\i}}

\def\mathbf{\vec}
\def\be{\begin{equation}}
\def\ee{\end{equation}}
\def\ba{\begin{eqnarray}}
\def\ea{\end{eqnarray}}

\def\bx
  {\hbox to.77778em{\hfil\vrule\vbox to.675em
  {\hrule width.6em\vfil\hrule}\vrule\hfil}}

\begin{document}
\begin{center}
{\Large\bf \boldmath Impact of eight quark interactions on chiral phase transitions II: Thermal effects} 

\vspace*{6mm}
{B. Hiller$^a$, A. A. Osipov$^{b,a}$, J. Moreira$^a$, A. H. Blin${^a}$}\\      
{\small \it $^a$ Departamento de F\ii sica, Universidade de Coimbra, P-3004-516 Coimbra, Portugal \\      
            $^b$ Laboratory of Nuclear Problems, JINR, Dubna, Russia}
\end{center}

\vspace*{6mm}

\begin{abstract}
In this talk attention is drawn to thermal properties due to the addition of eight quark interactions in the standard $SU(3)\times SU(3)$ chiral Nambu-Jona-Lasinio model (NJL) with 't Hooft interaction (NJLH). The schematic $SU(3)$ flavor limit with massless current quarks as well as the realistic
case $m_u=m_d\ne m_s$ are discussed.
\end{abstract}

\vspace*{6mm}
The extension of the NJLH model \cite{Nambu:1961}-\cite{Reinhardt:1988} to include 8$q$ (quark) interactions \cite{Osipov:2005b} finds its main motivation in the fact that it stabilizes the scalar effective potential of the theory. This has been discussed at length in \cite{Osipov:2005a},\cite{Osipov:2006a}. We show  here that they play an important role in the physics of chiral transitions at finite temperature \cite{OsipovT2:2008}.  
Before discussing the thermal properties, we give a summary of essential features of the model at $T=0$, relevant for the present discussion. 
One of the most attractive features of chiral multiquark interactions is the 
possibility of analyzing patterns of dynamical breakdown of chiral symmetry. In the original NJL model the symmetry is broken in the massless limit from a critical value of the 4$q$ interaction strength on, $G > G_{cr}$. The same happens if the $U_A(1)$ breaking 6$q$ lagrangian of 't Hooft is added with strength $\kappa$;  a realistic fit to the pseudoscalar and scalar mass spectra restricts $\kappa$ to the role of a perturbative effect on the spontaneously broken phase. This changes radically if the 8$q$ interactions are added, for which the most general spin 0 non-derivative combination consists of a sum of two terms (with strengths $g_1,g_2$), one of them ($\sim g_1$) violating the OZI-rule. Now, besides the former scenario, it is also possible to have the Wigner-Weyl phase (i.e. for $G < G_{cr}$ value) in coexistence with another minimum, induced by the higher multi-quark interactions. Realistic fits to low energy characteristics of the mesons show that the second minimum is induced by the strength $\kappa$. To these two possible symmetry breaking patterns one can assign the same meson mass spectra, (except for the $\sigma$ meson), due to an interplay of the 4$q$ and 8$q$ strengths. Turning now to the finite temperature case: if symmetry breakdown is induced by the 't Hooft interaction term, a substantial decrease of the transition temperature is achieved, bringing the model predictions closer to lattice results \cite{Aoki:2006}. This can be easily understood with help of fig. 1. On the left side the effective potential at $T=0$ shows coexistence of the Wigner-Weyl phase and symmetry broken phase, different curves correspond to different strength of $\kappa$. On the right the several curves of the effective potential represent different temperatures, calculated with model parameters fixed at $T=0$ (dashed curve, $G>G_{cr}$). 
Increasing the temperature the effective potential runs through configurations of the type shown on the left, changing curvature at origin. Thus starting already at $T=0$ from a configuration shown on the left will obviously reduce the transition temperature of the chiral transition. 
\begin{figure}[h]
\hspace{2cm}{\includegraphics[height=4cm]{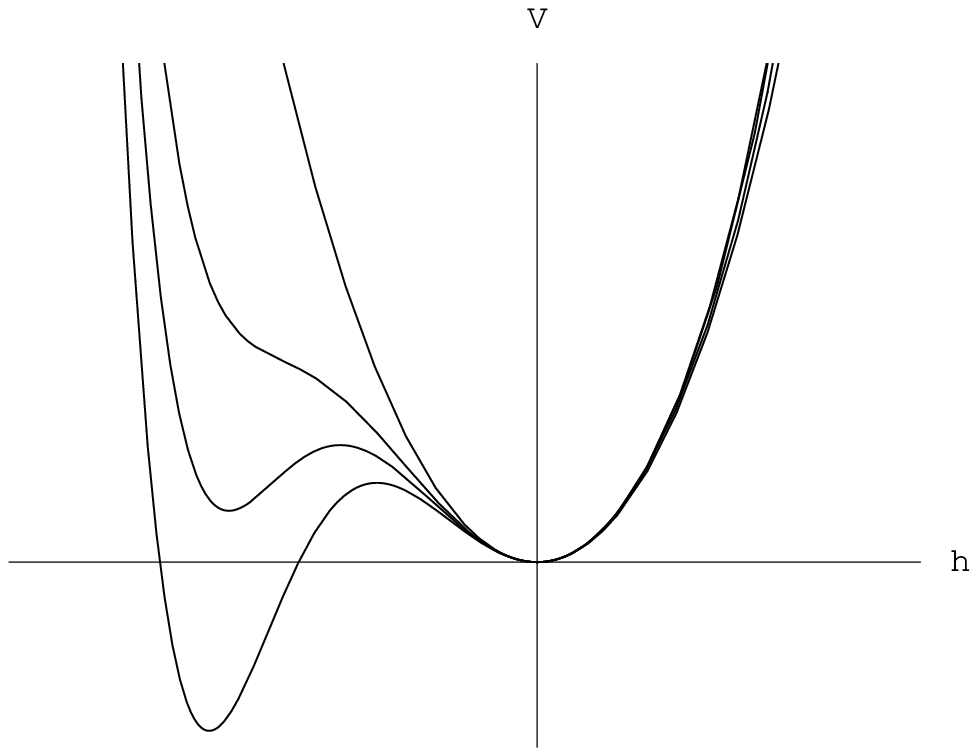}}  \hspace{2cm}{\includegraphics[height=4cm]{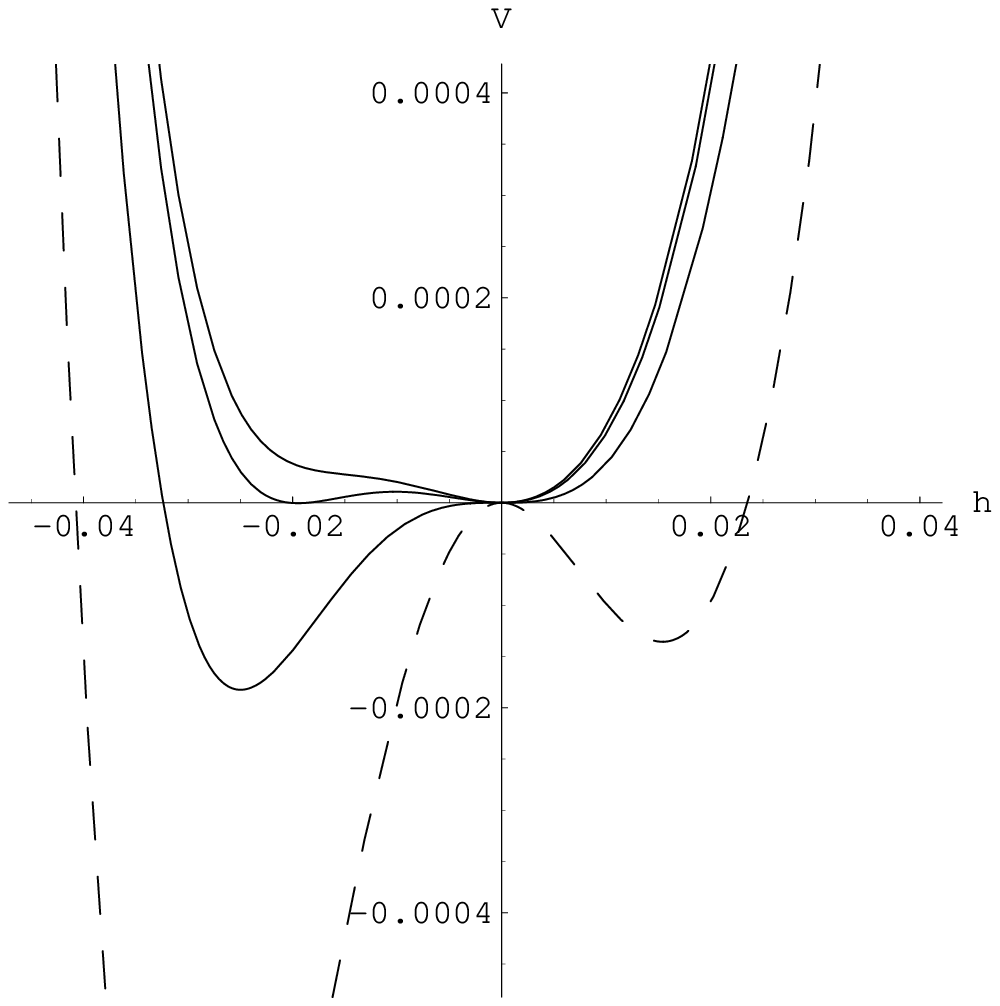}}  
\caption{Left: Effective potential at $T=0$ for different values of $\kappa$, axis through origin. Right: Effective potential at different $T$ values, with model parameters fixed at $T=0$ - dashed line. See also text. Figs. taken from \cite{Osipov:2006a},\cite{OsipovT2:2008} respectively.} 
\end{figure}
We discuss now the effect of realistic current quark masses. The gap equations at $T=0$ in the isospin limit $m_u=m_d\ne m_s$ are
\begin{equation}
\label{gap}
   \begin{array}{lcr}
   h_u+\displaystyle\frac{N_c}{6\pi^2} M_u
       \left(3I_0-\Delta_{us} I_1 \right)=0, 
   \hspace{1cm} h_s+\displaystyle\frac{N_c}{6\pi^2} M_s
       \left(3I_0+2\Delta_{us} I_1 \right)=0.
   \end{array}
\end{equation}
This system must be solved selfconsistently with the stationary phase
equations
\begin{equation}
\label{SPA}
   \left\{ \begin{array}{l}
\vspace{0.2cm}   
   Gh_u + \Delta_u +\displaystyle\frac{\kappa}{16}\, h_uh_s
   +\displaystyle\frac{g_1}{4}\, h_u(2 h_u^2+h_s^2)
   +\displaystyle\frac{g_2}{2}\, h_u^3=0, \\
\vspace{0.2cm}   
   Gh_s + \Delta_s +\displaystyle\frac{\kappa}{16}\, h_u^2
   +\displaystyle\frac{g_1}{4}\, h_s(2 h_u^2+h_s^2)
   +\displaystyle\frac{g_2}{2}\, h_s^3=0. 
   \end{array} \right.
\end{equation}
with
$\Delta_{us}=M_u^2-M_s^2$, $\Delta_l=M_l-m_l$, $h_l\sim$ condensates, $l=u,s$ for 
constituent quark masses $M_l$, $I_i = [2J_i(M_u^2)+J_i(M_s^2)]/3,$
$i=0,1,\ldots$. The  one-quark-loop integrals 
\begin{equation}
\label{ji}
   J_i(M^2)=\int\limits_0^\infty\frac{{\rm d}t}{t^{2-i}}\rho(t\Lambda^2) e^{-t M^2}, \hspace{1cm} \rho (t\Lambda^2)=1-(1+t\Lambda^2)\exp (-t \Lambda^2), 
   \end{equation}
where $\rho (t\Lambda^2)$ denotes the Pauli-Villars regularization kernel with two subtractions
and $\Lambda$ is an ultra-violet cutof\mbox{}f. 

To generalize to f\mbox{}inite temperatures, the quark loop integrals $J_0,J_1$
are modified, introducing the Matsubara frequencies
\begin{eqnarray}
\label{j0t}
   &&J_0(M^2)\rightarrow J_0(M^2,T) 
      =16\pi^2 T\!\sum_{n=-\infty}^{\infty} 
      \int\!\frac{{\rm d}^3{p}}{(2\pi )^3}
   \int\limits_0^\infty\!{\rm d}s\, \rho (s\Lambda^2) 
      e^{-s[(2n+1)^2\pi^2 T^2+\vec{p}^2+M^2]}.
\end{eqnarray}   
Using the Poisson formula
   $\sum_{n=-\infty}^{\infty} F(n) = \sum_{m=-\infty}^{\infty}
   \int_{-\infty}^{+\infty} {\rm d}x\, F(x) e^{i2\pi mx}$,
where $F(n)=\exp [-s(2n+1)^2\pi^2 T^2]$,
and after integration over 3-momentum $\vec{p}$ one obtains
\begin{eqnarray}
\label{j0t2}
   &&J_0(M^2,T)
   =\int\limits_0^\infty
             \frac{{\rm d}s}{s^2}\,\rho (s\Lambda^2) 
             e^{-s M^2} 
   \left[1+2\sum_{n=1}^{\infty} (-1)^{n}
              \exp\left(\frac{-n^2}{4sT^2}\right)\right],
\end{eqnarray}
and $ J_1 (M^2,T) = -\frac{\partial}{\partial M^2}\, J_0 (M^2,T). $
At $T=0$ the mode $n=0$ decouples from $T\ne 0$ modes $n>0$. One 
recovers the covariant expression at $T=0$ and 
$\lim_{T\to\infty}J_{0,1}(M^2,T)=0.$

In Fig. 2 we display solutions $M_l(T)$, $l=u,s$ of (\ref{gap})-(\ref{SPA}) to the gap equations for the $m_u=m_d\ne m_s$ case, for the parameter sets of \cite{OsipovT2:2008}. Depending on the strength $g_1$ of the 8$q$ interactions we obtain either a crossover transition (left) or a first order transition (middle); units MeV. The three sets of $M_u,M_s$ are shown as solid line, dotted line, dashed line, and start at $T=0$ as minima, maxima and saddle points of the effective potential, respectively. On the right side (units GeV) we show the low pseudoscalar and scalar meson spectra for the crossover case obtained    
with parameters $G, \kappa , g_1, g_2, m_l, \Lambda$ independent on
temperature. 

\begin{figure}[h]
{\includegraphics[height=3.2cm]{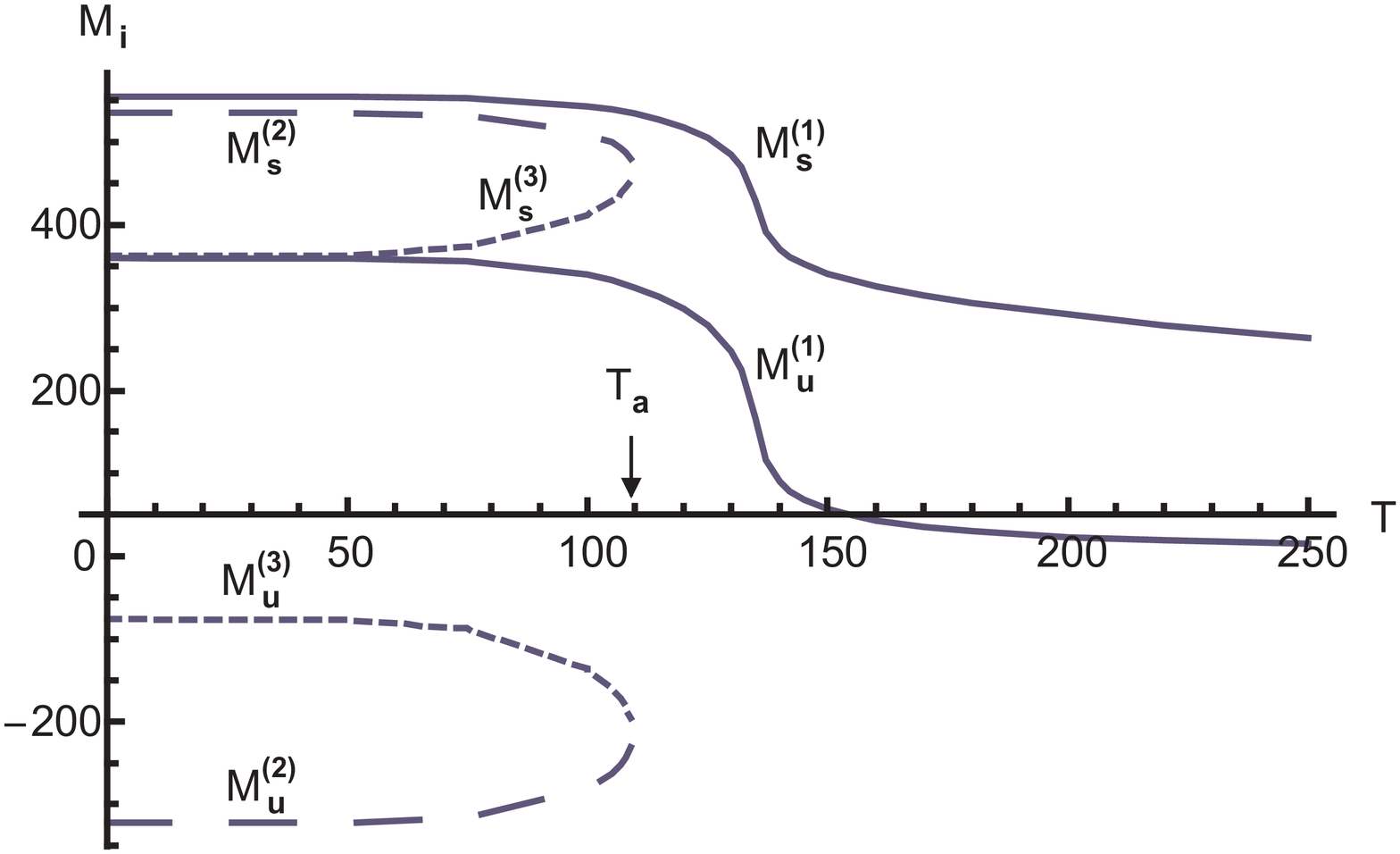}}\hspace{0.3cm}{\includegraphics[height=3.2cm]{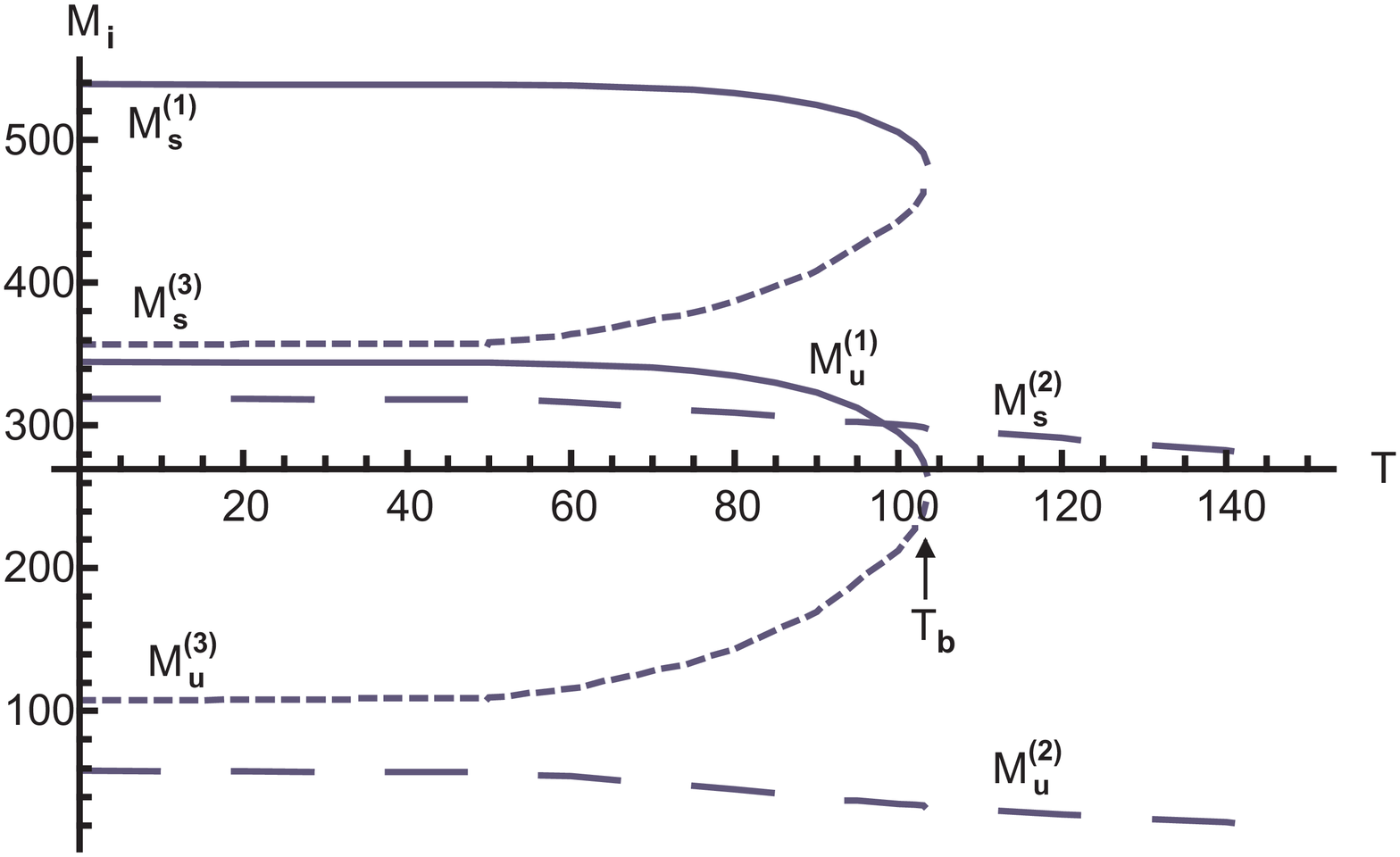}}\hspace{.5cm}{\includegraphics[height=3.5cm]{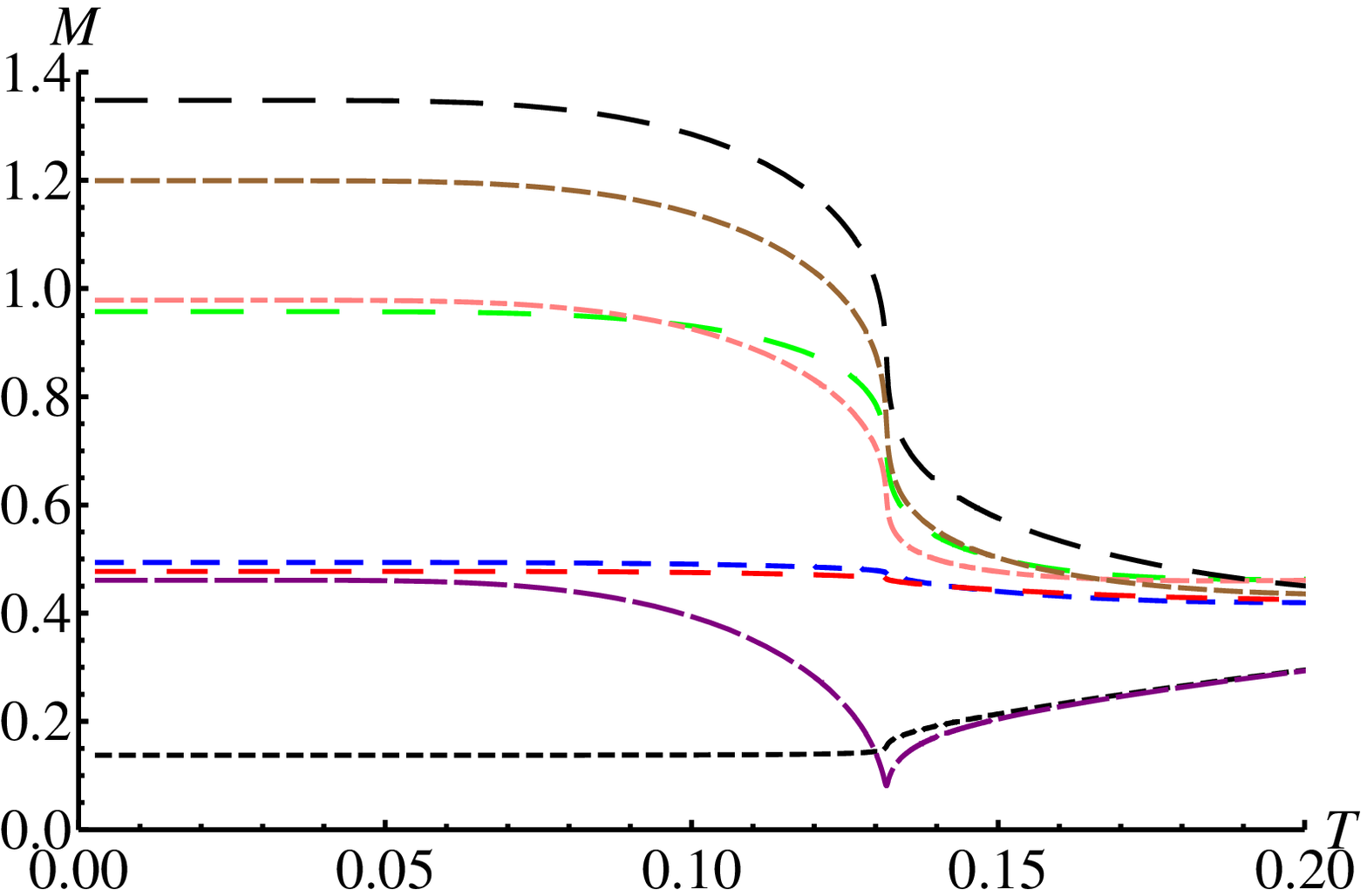}}
\caption{solutions to the gap equations for the light $M_u$ and strange quarks $M_s$; left: crossover case (only one set of solutions is always positive valued), middle: 1st order transition (all sets of solutions are positive valued); right: pseudoscalar and scalar mass spectrum for the crossover case ($\pi,\sigma,\eta,K,\eta',K_0^*,a_0,f_0(980)$, bottom to top). Details: text and \cite{OsipovT2:2008}. } 
\end{figure}


In conclusion, the chiral eight quark interactions, which cure the
global instability of the $4q+6q$ vacuum, play also a relevant role at finite $T$, acting as a chiral thermometer for chiral transitions. Temperature, slope and nature of transition are regulated by their strength $g_1$, which can be adjusted in consonance with $G$, with all other parameters frozen and leaving meson mass spectra at $T=0$ unaffected, except for $m_\sigma$. Finally, they allow to start from configurations which at $T=0$ have dynamical chiral symmetry breaking induced by the strength of $\kappa$ of the 't Hooft $6q$ terms, which turn out to be more favorable to a reduction of the transition temperature, compared to the conventual $4q$ strength $G$.

This work has been supported in part by grants provided by 
FCT: POCI/2010, FEDER,
POCI/FP/63930/2005, POCI/FP/81926/2007 and SFRH/BD/13528/2003.









\end{document}